\documentstyle[12pt,epsfig,cite]{article}

\newlength{\dinwidth}
\newlength{\dinmargin}
\def\docnum#1{\hbox to \hsize{\hskip123mm\hbox{#1}\hss}}
\def\date#1{\edef\@temp{#1}\ifx\@temp\@empty\def\@temp{\today}\fi
\hbox to \hsize{\hskip123mm\hbox{\@temp}\hss}}
\def\title#1{\vskip 0.8in plus 2in\begin{center}%
{\Large\bf#1\par}\vskip1.5em\end{center}\vskip 1in}
\def\@makefnmark{\hbox{$^{\@thefnmark)}$}}
\def\author#1{%% Treat the list of authors
\setcounter{footnote}{0}\def\@currentlabel{}%
\begingroup\def\thefootnote{\arabic{footnote}}
\def\@makefnmark{\hbox{$^{\@thefnmark)}$}}
\global\@topnum\z@ \large\begin{center}{\lineskip.5em
\begin{tabular}[t]{c}#1\end{tabular}\par}
\end{center}\par\vskip1.5em\@thanks\endgroup}

\def\abstract{\vskip0.8in plus 3in\begin{center}{\large\bf Abstract}\end{center}\quotation}

\newcommand{\QG}   {{\bf{Q}}}
\newcommand{\QGz}  {{\bf{Q}}^0}
\newcommand{\QGzi} {{\bf{Q}}_i^0}

\newcommand{\PG}   {{\bf \phi}}

\newcommand{\ppb}  {$\rm{p\bar{p}}\;$}

\newcommand{\ee}   {e$^+$e$^-$}

\newcommand{\qj}   {{\bf{q}}_j}

\newcommand{\zum}  {{\rm{\Sigma}}}
\newcommand{\dint} {{\rm{d}}}

\newcommand{\E}    {{\rm{e}}}
\newcommand{\I}    {{\rm{i}}}
\newcommand{\muv}  {{\bf{\mu}}}
\newcommand{\ppbf} {${\bf{p\bar{p}}}\;$}

\begin{document}

\begin{titlepage}
\flushright{DFF 284/07/1997}
\flushright{July 1997}
\title{Thermal hadron production in high energy collisions} 
\vspace{-2.0cm}
\centerline{\large{F. Becattini}} 
\vspace{0.5cm} 
\centerline{\it{INFN Sezione di Firenze}}  
\centerline{\it{Largo E. Fermi 2, I-50125 Firenze, ITALY}} 
\centerline{e-mail: becattini@fi.infn.it}

\begin{abstract}
 It is shown that hadron abundances in high energy \ee, pp and 
 \ppb collisions, calculated by assuming that particles originate 
 in hadron gas fireballs at thermal and partial chemical equilibrium, 
 are in very good agreement with the data. 
 The freeze-out temperature of the hadron gas fireballs turns out 
 to be nearly constant over a large center of mass energy range and
 not dependent on the initial colliding system. The only deviation
 from chemical equilibrium resides in the incomplete strangeness 
 phase space saturation. Preliminary results of an analysis of 
 hadron abundances in S+S and S+Ag heavy ion collisions are 
 presented.

\end{abstract}
\vspace*{1.5cm}
\centerline{\it{Talk given at the "Strangeness and Quark Matter 97" conference}}
\centerline{\it{April 14-19 1997, Santorini (Greece) (to be published in the Proceedings)}}

\end{titlepage}

\section{Introduction}
 
 The average multiplicities of particles produced in high energy 
 collisions are very useful tools to investigate the process of
 hadron production in virtue of some peculiar features. Unlike 
 momentum spectra, hadron abundances (and correlations) are 
 Lorentz-invariant quantities; hence they do not depend on 
 complicated collective motions possibly present in the system 
 and may be calculated in the local comoving frames. In elementary 
 collisions, such as \ee, they are a direct and unique probe of 
 the hadronization process since they are independent of the 
 perturbative parton dynamics which is inherited by hadrons mainly
 in their momentum spectrum. Therefore, it is very important to study 
 hadron abundances in order to reveal the basic mechanisms governing 
 hadron production in all kinds of collisions.\\
 In the following sections we will sum up briefly the 
 statistical-thermodynamical approach to the problem of hadron production 
 and we will show its stunning capability of fitting all existing hadron 
 average multiplicities data in \ee, pp and \ppb collisions by using 
 only three free parameters.
 A preliminary analysis of hadron abundances measured in heavy 
 ion collisions in full phase space within the same model will
 be discussed in Sect. 4.   
   
\section{The model}

 The thermodynamical model of hadron production in \ee, pp, \ppb has
 been described in detail elsewhere \cite{beca,erix,behe}; in this 
 section it is briefly summarized.\\
 The basic assumption of the model is the formation of an arbitrary 
 number of hadron gas fireballs moving away from the primary interaction 
 region each with its own collective momentum. The parameters describing
 the $i^{th}$ hadron gas fireball at thermal and chemical equilibrium 
 are the temperature $T_i$ and the volume $V_i$ in {\em its rest frame} 
 as well as its quantum numbers electric charge $Q$, baryon number $N$,
 strangeness $S$, charm $C$ and beauty $B$. The partition function of 
 this system is calculated in the framework of the canonical formalism 
 of statistical mechanics, namely by summing only over the multi-hadronic
 states having the same quantum numbers of the fireball. Therefore, 
 if $\QGzi=(Q,N,S,C,B)$ is the vector of fireballs quantum numbers 
 and $\QG$ is the vector of quantum numbers of a particular multi-hadronic 
 state, the partition function of the fireball reads:
 
\begin{equation}
  Z(\QGzi) = \sum_{\rm{states}} \E^{-E/T_i} \delta_{\QG,\QGzi} \; .
\end{equation}
 A parameter $\gamma_s$ accounting for a possibly incomplete strangeness
 chemical equilibrium is introduced in the partition function by multiplying 
 by $\gamma_s^{s}$ the Boltzmann factors $\E^{-\epsilon_j/T}$ associated 
 to the $j^{th}$ hadron where $s$ is the number of its valence strange quarks 
 and anti-quarks.\\
 The average multiplicity of any hadron species in the $i^{th}$ fireball
 can be derived from the partition function (1). As this quantity
 depends on the quantum vector $\QGzi$, the overall average multiplicity 
 depends on the number of fireballs $N$ and on their quantum configuration 
 $\{\QG_1^0,\ldots,\QG_N^0\}$. In principle any configuration may occur 
 provided that $\sum_{i=1}^N \QGzi = \QGz$, where $\QGz$ is the quantum
 vector fixed by the initial state. However, it can be shown \cite{behe}
 that the overall average multiplicity of any hadron indeed depends only on the
 global quantities $\QGz$ and $V=\sum_{i=1}^N V_i$ (namely the sum of 
 all fireball volumes in their rest frames), provided that the temperatures 
 and the strangeness suppression factors $\gamma_s$ are the same for all    
 fireballs and the probabilities $w(\QG_1^0,\ldots,\QG_N^0)$ of occurrence 
 of a given quantum configuration are chosen to be:
 
\begin{equation}
     w(\QG_1^0,\ldots,\QG_N^0) = \frac{\delta_{\zum_i \QGzi,\QGz} \prod_{i=1}^N 
      Z_i(\QGzi)}{\sum_{\QG_1^0,\ldots,\QG_N^0} \!\!\! \delta_{\zum_i \QGzi,\QGz} 
      \prod_{i=1}^N Z_i(\QGzi)}  .    
\end{equation}
 It can be proved that this choice corresponds to the minimal deviation of 
 the system from global (i.e. thermal, chemical and mechanical) equilibrium.
 After making use of the probabilities (2) to average the hadron production over
 all possible quantum configurations, the overall average multiplicity of 
 the $j^{th}$ hadron turns out to be:
 
\begin{eqnarray}
   \langle\!\langle n_j \rangle\!\rangle &=&  \frac{1}{(2\pi)^5}
    \int \dint^5 \phi \,\, \E^{\,\I\, \QGz \cdot \phi} \exp  
    [ V \sum_j F_j(T,\gamma_s,\PG)] 
    \nonumber \\ 
   &\times& \frac {(2J_j+1)\,V}{(2\pi)^3} \int 
      \frac {\dint^3 p}{\gamma_s^{-s_j}
      \exp \,(\sqrt{p^2+m^2_j}/T+\I \qj \cdot \PG) \pm 1} \; , 
\end{eqnarray}   
 where the upper sign is for fermions, the lower for bosons and:

\begin{equation}  
 F_j(T,\gamma_s,\PG)=\sum_j \frac{(2J_j+1)V}{(2\pi)^3} \int \dint^3 p \,\, 
   \log \, (1 \pm \gamma_s^{s_j} \E^{-\sqrt{p^2+m_j^2}/T - 
   \I \qj \cdot \phi})^{\pm 1} \; .
\end{equation}
 Thus, under the previous assumptions, the hadron yields (3) depend only 
 on three unknown parameters $T$, $\gamma_s$ and $V$; the latter re-absorbs 
 the dependence on the number of fireballs. These unknown parameters have 
 to be determined by fitting the calculated multiplicities to the measured 
 ones at each center of mass energy.  
  
\section{Results in \ee, pp and \ppbf collisions}

 In order to calculate hadron abundances to be compared with experimental 
 data the primary yield of each hadron species calculated with eq.~(4)
 is added to the contribution stemming from the decay of heavier hadrons, 
 which is calculated by using experimentally known decay modes and 
 branching ratios \cite{pdg,jet}. All light-flavored hadrons up to a 
 mass of 1.7 GeV and all heavy-flavored states inserted in the JETSET 
 tables \cite{jet} have been used as primary species. The effect of this
 cut-off of hadron mass spectrum on final results has been shown to be negligible
 \cite{beca,behe}.\\
 The primary yield of resonances has been determined by convoluting the eq.~(4)
 with a relativistic Breit-Wigner function within $2\Gamma$ from the central
 mass value.\\   
 The measurements from different experiments have been averaged according to 
 a procedure described in ref. \cite{dean} taking into account {\it a posteriori} 
 disagreements and correlations.\\
 Since the temperature is expected to be ${\cal O}(100)$ MeV the thermal
 production of heavy flavored hadrons can be neglected while the perturbative
 production is significant only in \ee collisions, where c and b quarks are
 created in the primary interaction and do not re-annihilate. In this case
 the presence of one charmed (bottomed) flavored hadron-anti-hadron pair is 
 demanded in a fraction of events $\sigma({\rm e}^+{\rm e}^- \rightarrow {\rm c} \overline
 {\rm c} ( {\rm b} \overline {\rm b}))/\sigma({\rm e}^+{\rm e}^- \rightarrow 
 {\rm hadrons})$.\\
 The fit is performed by minimizing the $\chi^2$:
 
\begin{equation}
  \chi^2 = \sum_i \frac{(n_i[{\rm theo}] - n_i[{\rm expe}])^2}{\sigma_i^2}
\end{equation}
 as a function of $T$, $V$ and $\gamma_s$. The errors $\sigma_i$ include
 contributions from uncertainties on masses, widths and branching
 ratios of various hadrons involved in the decay chain process; they have been
 determined with an iterative fit procedure \cite{beca,behe}.
\begin{table*}[htb]
 \caption[]{Values of fitted parameters. The parameter $V T^3$ 
  has been used instead of $V$ in hadronic collisions because less 
  correlated to the temperature. The additional errors within brackets have 
  been estimated by excluding data points deviating the most from fitted 
  values and repeating the fit.}
   \begin{center}
   \begin{tabular}{| c || c | c | c | c |}
        \hline
         \multicolumn{5}{c}{\ee collisions} \\ \hline
   $\sqrt s$(GeV)&  Temp. (MeV)        & Volume(Fm$^3$)  & $\gamma_s$             &  $\chi^2$/dof  \\ \hline
    29           & $163.6\pm3.6 $      & $26.7\pm4.1 $   & $0.724\pm0.045$        &  24.7/13    \\ 
    35           & $165.2\pm4.4 $      & $24.9\pm4.7 $   & $0.788\pm0.045$        &  10.5/8     \\ 
    44           & $169.6\pm9.5 $      & $23.2\pm8.7 $   & $0.730\pm0.060$        &  4.9/4      \\ 
    91           & $160.3\pm1.7(3.3)$  & $50.0\pm3.9 $   & $0.673\pm0.020(0.028)$ &  70.1/22    \\ \hline
       \multicolumn{5}{c}{pp collisions} \\ \hline
 $\sqrt s$(GeV)  &   Temp. (MeV)      & $V T^3$          & $\gamma_s$             &  $\chi^2$/dof \\ \hline
    19.4         & $190.8\pm27.4$     & $5.8\pm3.1$      & $0.463\pm0.037$        &  6.4/4    \\ 
    23.6         & $194.4\pm17.3$     & $6.3\pm2.5$      & $0.460\pm0.067$        &  2.4/2    \\ 
    26.0         & $159.0\pm9.5$      & $13.4\pm2.7$     & $0.570\pm0.030$        &  1.9/2     \\ 
    27.4         & $169.0\pm2.1(3.4)$ & $11.0\pm0.69$    & $0.510\pm0.011(0.025)$ &  136.4/27  \\ \hline
       \multicolumn{5}{c}{\ppb collisions} \\ \hline
 $\sqrt s$(GeV)  &   Temp. (MeV)      & $V T^3$          & $\gamma_s$             &  $\chi^2$/dof \\ \hline    
    200          & $175.0\pm14.8$     & $24.3\pm7.9$     & $0.537\pm0.066$        &  0.70/2    \\ 
    546          & $181.7\pm17.7$     & $28.5\pm10.4$    & $0.557\pm0.051$        &  3.78/1    \\ 
    900          & $170.2\pm11.8$     & $43.2\pm11.8$    & $0.578\pm0.063$        &  1.8/2    \\ 
      \hline
   \end{tabular}
\end{center}
\end{table*}
 The results of the fit are shown in table 1. The quoted numbers are the same
 as in refs. \cite{erix,behe} except at $\sqrt s =91.2$ GeV where the
 fit has been repeated with new LEP measurements \cite{newlep} (see fig. 1).\\
 The fit quality is remarkably good at all center of mass energy points. 
 The most interesting result is undoubtedly the uniformity, within the 
 fit errors, of the freeze-out 
 temperature values independently of kind of reaction and center of mass energy. 
 The fact that $\gamma_s$ is always less than 1 demonstrates that strangeness 
 chemical equilibrium is not reached in any of the examined collisions. 
 Nevertheless, it is worth noticing that $\gamma_s$ is higher in \ee collisions 
 than in hadronic collisions at the same center of mass energy.\\
 The use of the canonical formalism is essential since the system turns out
 to be small enough to generate charged hadron ($\qj \ne 0$) suppression with 
 respect to neutral ones ($\qj =0$) even in an initially neutral system ($\QGz = 0$);
 for a more detailed discussion see refs. \cite{erix,behe}.\\  
 All the fits have been performed by using as experimental input the measured 
 yields of light-flavored hadrons. Once the parameters of the model are 
 determined it is possible to predict the heavy-flavored hadrons abundances
 provided that the production rate of c and b quark pairs is known. 
 In table 2 predictions for $\sqrt s = 91.2 $ GeV are compared to actual
 LEP experiments measurements \cite{hf} averaged according to the procedure 
 mentioned above; the agreement is indeed very good.  
   
\begin{table*}[htb]
 \caption[]{Predictions of heavy flavored hadron abundances at $\sqrt s =91.2$
  GeV obtained by using $T$, $V$ and $\gamma_s$ parameters quoted in table 1 and
  $R_c =0.17$, $R_b =0.22$ according to LEP measurements \cite{rbb}. The B$_s^{**}$
  prediction is affected by the interpretation of the observed peaks as four different
  states or two different states (within brackets).}

   \begin{center}
   \begin{tabular}{| c || c | c | c |}
      \hline 
 {\bf Hadron}                   & Prediction    &  Measured             &  Residual      \\ \hline
D$^+$                           &  0.0926       &  0.087$\pm$0.008      & -0.67   \\ 
D$^0$                           &  0.233        &  0.227$\pm$0.012      & -0.50   \\ 
D$_s$                           &  0.0579       &  0.066$\pm$0.010      & +0.81   \\ 
D$^{*+}$                        &  0.108        &  0.0880$\pm$0.0054    & -3.7     \\ 
D$_s^+$/c-jet                   &  0.103        &  0.128$\pm$0.027      & +0.92    \\ 
D$_1$/c-jet                     &  0.0347       &  0.038$\pm$0.009      & +0.37   \\ 
D$^*_2$/c-jet                   &  0.0471       &  0.135$\pm$0.052      & +1.7     \\  
D$_{s1}$/c-jet                  &  0.00536      &  0.016$\pm$0.0058     & +1.8     \\  
B$^0$/b-jet                     &  0.412        &  0.384$\pm$0.026      & -1.1     \\ 
B$^*$/B                         &  0.692        &  0.747$\pm$0.067      & +0.82   \\ 
B$^*$/b-jet                     &  0.642        &  0.65 $\pm$0.06       & +0.13   \\ 
B$_s$/b-jet                     &  0.106        &  0.122$\pm$0.031      & +0.52   \\ 
B$^{**}_{u,d}$/b-jet            &  0.206        &  0.26 $\pm$0.05       & +1.0    \\ 
B$^{**}$/B                      &  0.251        &  0.27 $\pm$0.06       & +0.32   \\ 
B$^{**}_s$/b-jet                & 0.021(0.011)  &  0.048$\pm$0.017      & +1.6    \\ 
B$^{**0}_s$/B$^+$               & 0.026(0.013)  &  0.052$\pm$0.016      & +1.6     \\ 
$\Lambda_c^+$                   &  0.0248       &  0.0395$\pm$0.0084    & +1.7   \\ 
b-baryon/b-jet                  &  0.0717       &  0.115 $\pm$0.040     & +1.1    \\ 
$(\Sigma_b+\Sigma^*_b)$/b-jet   &  0.0404       &  0.048 $\pm$0.016     & +0.48   \\
$\Sigma_b/(\Sigma_b^*+\Sigma_b)$&  0.411        &  0.24  $\pm$0.12      & -1.4     \\ 
      \hline
   \end{tabular}
\end{center}
\end{table*}

\section{Thermal fits in heavy ion collisions}     
      
 The model described in Sect. 2 may be used to fit hadron abundances
 measured in heavy ion collisions, provided that the same assumptions still 
 hold. Comparison of thermal calculations with experimental data have been 
 done recently by several authors with a grand-canonical, rather than canonical, 
 approach and by using multiplicities measured either in a restricted rapidity 
 range or in full phase space \cite{soll,munz,satz,spiel}.\\ 
 In principle the canonical formalism is the only correct one in that it 
 ensures the exact conservation of initial quantum numbers. However, if
 the volume $V$ is very large, it can be shown (see ref. \cite{behe}) that 
 the formula (3) giving the average primary $j^{th}$ hadron multiplicity 
 in the canonical formalism reduces to: 
 
 \begin{eqnarray}
  \langle\!\langle n_j \rangle\!\rangle &=&  (2J_j+1) \, \frac{V}{(2\pi)^3} \,
    \sum_{n=1}^{\infty} (\pm 1)^{n+1} \gamma_s^{ns_j}  \nonumber \\
  &&  \times \int \dint^3 p \,\, \E^{-n \sqrt{p^2+m_j^2}/T} \,\, 
     \E^{n \QG {\sf{A}}^{-1} \qj/2} \,\, \E^{-n^2 \qj {\sf{A}}^{-1} \qj/4} \; ,
 \end{eqnarray}
 by using a saddle-point approximation of the $\phi$-integrals in eq.~(3). 
 ${\sf{A}}$ is a $N\times N$ matrix, where $N$ is the dimension of the quantum
 vectors $\QG,\qj$, whose elements are proportional to $V$. 
 Hence, in the limit $V \rightarrow \infty$, the second exponential factor 
 in the above equation goes to 1 as the $\qj$ terms are finite (i.e. the 
 hadrons quantum numbers). On the other hand, the first exponential factor 
 can be written $\exp[n \muv \cdot \qj]$ where $\muv$ is a set of $N$ traditional 
 chemical potentials; the grand-canonical formalism is recovered in the large volume 
 limit. In heavy ion collisions one expects the canonical factor
 $\exp[-n^2 \qj {\sf{A}}^{-1} \qj/4]$ to be a small correction of the grand
 canonical formulae as the particle multiplicities, hence the volume, are 
 very large compared to pp or \ee collisions.\\
 We fitted hadron abundances measured in SS \cite{ss} and SAg \cite{sag} 
 collisions in full phase space by using four free parameters: $T$, $V$, 
 $\gamma_s$ and $\mu_b$, the baryochemical potential. The strangeness and
 electric chemical potential $\mu_s$ and $\mu_q$ have been determined with
 the constraints of strangeness neutrality and conservation of the initial
 electric charge/baryon number initial ratio:
 
\begin{eqnarray}
 && \sum_j S_j \langle\!\langle n_j \rangle\!\rangle = 0 \nonumber \\ 
 && \sum_j Q_j \langle\!\langle n_j \rangle\!\rangle = \frac{Z}{A} 
    \sum_j N_j \langle\!\langle n_j \rangle\!\rangle  \; .
\end{eqnarray}
 The results of the fit are shown in table 3 while the comparison between
 fitted and experimental average multiplicities are shown in table 4. Due
 to the strong correlation between $T$ and $V$ we chose to fit the parameter 
 $VT^3 \exp[-0.7 \, {\rm GeV}/T]$ instead of $V$.
\begin{table*}[htb]
 \caption[]{Values of fitted parameters in SS and SAg collisions. Also
            quoted the calculated chemical potentials $\mu_s$ and $\mu_q$.}
   \begin{center}
   \begin{tabular}{| c || c | c |}
        \hline
     Parameter              &  SS            & SAg             \\ \hline
       $T$ (MeV)            & 182.1$\pm$9.0  & 180.0$\pm$3.2   \\
$VT^3\exp[-0.7 {\rm GeV}/T]$& 3.51$\pm$0.14  & 5.43$\pm$0.35   \\  
      $\gamma_s$            &0.732$\pm$0.037 & 0.830$\pm$0.061 \\  
      $\mu_b/T$             &1.243$\pm$0.071 & 1.323$\pm$0.069 \\
      $\chi^2/$dof          &  17.2/5        &  5.5/3          \\ \hline
      $\mu_s/T$             &  -0.332        & -0.364          \\
      $\mu_q/T$             &  -0.0222       & -0.00316        \\ \hline           
   \end{tabular}
\end{center}
\end{table*}
\begin{table*}[htb]
 \caption[]{Comparison between fitted and measured multiplicities in SS and
            SAg collisions.}
   \begin{center}
   \begin{tabular}{| c || c | c | c |}
      \hline 
 {\bf Particles SS}   &  Fitted  &  Measured	    &  Residual  \\ \hline
  Baryons-Antibaryons &  54.57   &    54$\pm$3      & -0.19	  \\
  $h^-$               &  93.41   &    98$\pm$3      & +1.53      \\
  K$^+$               &  12.61   &   12.5$\pm$0.4   & -0.28	  \\
  K$^-$               &  7.456   &    6.9$\pm$0.4   & -1.39	  \\
  K$^0_s$             &  9.834	 &   10.5$\pm$1.7   & +0.39	  \\
 $\Lambda$            &  7.798   &    9.4$\pm$1.0   & +1.60	  \\
 $\bar \Lambda$       &  1.425   &    2.2$\pm$0.4   & +1.94	  \\
 p - $\bar{\rm p}$    &  22.59   &   21.2$\pm$1.3   & -1.07	  \\
  $\bar{\rm p}$       &  2.094   &    1.15$\pm$0.4  & -2.36	  \\ \hline \hline 
 {\bf Particles SAg}  &  Fitted  &  Measured	    &  Residual  \\ \hline
  Baryons-Antibaryons &  92.02   &    90$\pm$9      & -0.22      \\
  $h^-$               &  152.04  &   160$\pm$8      & +1.00      \\
  K$^0_s$             &  17.49   &  15.5$\pm$1.5    & -1.33	 \\
 $\Lambda$            &  14.39   &  15.2$\pm$1.2    & +0.68	 \\
 $\bar \Lambda$       &  2.440   &   2.6$\pm$0.3    & +0.53	 \\
 p - $\bar{\rm p}$    &  36.76   &    34$\pm$4      & -0.68	 \\
  $\bar{\rm p}$       &  3.043   &   2.0$\pm$0.8    & -1.31	 \\ \hline 
   \end{tabular}
\end{center}
\end{table*} 
 It should be mentioned that
 these results have been obtained by using only the experimental errors 
 without taking into account the uncertainties arising from hadron parameters 
 like masses, widths and branching ratios. \\
 The resulting elements of the ${\sf{A}}$ matrix range between -0.02 and 
 0.06 in SS collisions and between -0.012 and 0.039 in SAg, confirming the 
 proximity to the grand-canonical regime.\\
 The fitted temperature is compatible with that found in \ee, pp, and \ppb 
 collisions and the quality of the fit is good as well. 
 Strangeness chemical 
 equilibrium is not reached as demonstrated by the $\gamma_s$ values $<1$
 although there is a clear increase with respect to pp and \ppb collisions.
 Our results differ from those obtained in ref. \cite{soll} mainly because 
 of the available larger number of data points and the use of updated hadron 
 parameters in the decay chain. 
 A new fit performed by one of the authors of ref. 
 \cite{soll} shows a clear consistency with our results \cite{soll2}.

\section{Conclusions}
 
 The analysis of hadron abundances in \ee, pp and \ppb collisions performed
 in a suitable canonical formalism is in very good agreement with the hypothesis 
 of {\em local} thermal and chemical equilibrium. The most interesting results 
 of the thermal fits to experimental data is the constant value of freeze-out 
 temperature in all three kinds of collisions independently of center of mass 
 energy. This fact indicates that the transition from quarks-gluons to hadrons 
 occurs in a purely statistical fashion at critical values of pre-hadronic matter 
 parameters (such as energy density or pressure) corresponding to a (partially) 
 equilibrated hadron gas at $T_c \simeq 170$ MeV. Furthermore, evidence is found 
 for an incomplete strangeness phase space saturation.\\
 The preliminary analysis of hadron abundances in full phase space in 
 SS and SAg heavy ion collisions resulted in a good agreement with the data
 as well and a temperature value consistent with that found in elementary 
 collisions.\\
 The strangeness enhancement going from pp to heavy ion collisions is explained  
 by two different effects: the increase of the extension of the system 
 reduces 
 the suppression due to strangeness conservation (canonical suppression) whilst
 the increase of $\gamma_s$ further raises the yield of particles containing
 strange quarks.     

\section*{Acknowledgements}

 I wish to express my gratitude to M. Gazdzicki who provided me with 
 heavy ion collisions data and for many illuminating discussions 
 about their analysis. I warmly thank J. Cleymans, U. Heinz, H. Satz,
 and J. Sollfrank for useful and stimulating discussions. I would like to 
 thank all the participants and the organizers of the conference for having provided 
 a favorable and friendly work atmosphere and for having chosen a wonderful
 environment.

\newpage

\begin{figure}[htbp]
\mbox{\epsfig{file=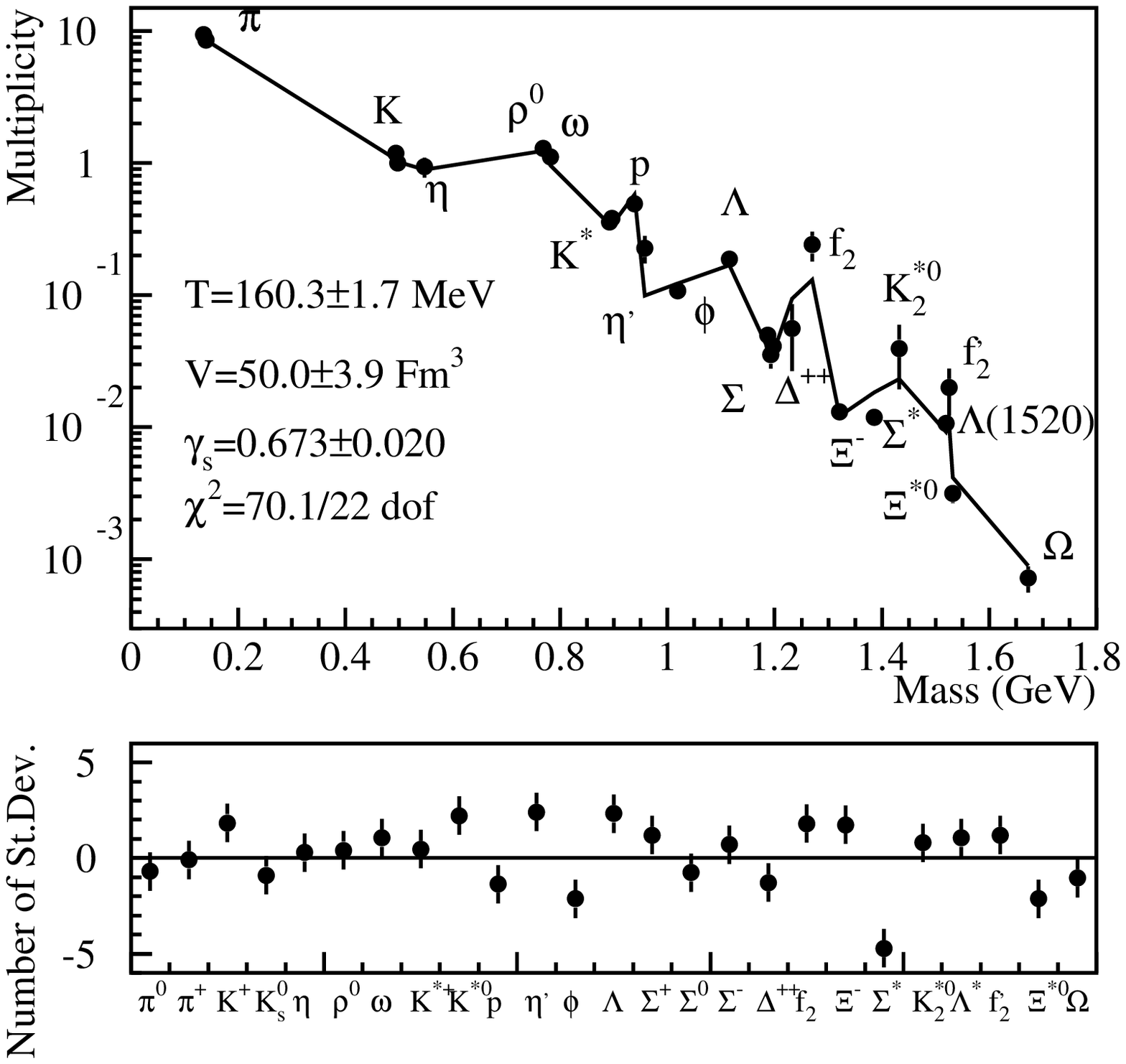,width=16cm}}
\caption{Fit of hadron average multiplicities at $\sqrt s =$ 91.2 GeV
 measured by LEP experiments. Above: black dots are the experimental
 data, the solid line connects fitted values. 
 Below: residuals distribution.}
\end{figure}

\end{document}